# Magnetocaloric effect and spin-phonon correlations in $RFe_{0.5}Cr_{0.5}O_3$ (R = Er and Yb) compounds

Kavita Yadav, Gurpreet Kaur, Mohit K. Sharma, and K. Mukherjee

School of Basic Sciences, Indian Institute of Technology Mandi, Mandi 175005, Himachal Pradesh, India

**Abstract**

We report the results of our investigation of the physical properties of mixed metal oxides $RFe_{0.5}Cr_{0.5}O_3$ (R = Er and Yb). $ErFe_{0.5}Cr_{0.5}O_3$ undergoes an antiferromagnetic ordering around 270 K followed by spin reorientation (SR) transitions around 150 and 8 K respectively. In contrast, in $YbFe_{0.5}Cr_{0.5}O_3$ a single SR transition is noted at 36 K, below the AFM ordering temperature of 280 K. In $ErFe_{0.5}Cr_{0.5}O_3$, a significant value of magnetic entropy change ($\Delta S_M$) ~ -12.4 J/kg-K is noted near the 2$^{nd}$ SR transition, however, this value is suppressed in $YbFe_{0.5}Cr_{0.5}O_3$. Temperature dependent dielectric permittivity of $ErFe_{0.5}Cr_{0.5}O_3$ and $YbFe_{0.5}Cr_{0.5}O_3$ at different frequencies, reveal the presence of Debye-like relaxation behaviour in both compounds, which can be due to the effect of charge carrier hopping between localized states of Fe and Cr ions. Temperature dependent Raman scattering studies divulge that spin-phonon coupling plays a crucial role in defining the physical properties of these compounds.

## 1. Introduction

In the past few decades, mixed metal oxides of the form RB′$_{1-x}$B″$_x$O$_3$ (R = rare-earth ions, B = 3$d$/4$d$/5$d$ transition metal ions) have received considerable attention due to novel physical properties like magnetization reversal, large magnetocaloric effect (MCE), multiferrocity exhibited by them [1-6]. Due to these functional properties, they can be useful in the area of technological applications, for example, in magnetic read heads for hard drives, sensors, spintronics, magnetic refrigeration technology [7-11]. Reports on rare-earth based oxides also suggest that they exhibit insulating behaviour which makes them potential candidates for electrical energy storage devices [12]. In these types of oxides, structural, electrical and magnetic properties are strongly correlated, and these properties can be tuned by the change in the R-O, B′-O and B″-O bond lengths. This change can be quantified by tolerance factor ($t$) which is expressed as $t = (r_R+r_O)/\sqrt{2}\,(((r_{B'}+r_{B''})/2)+r_O)$, where $r_R$, $r_{B'}$, $r_{B''}$ and $r_O$, are the ionic radii of the elements, and oxygen atom at R, B′, B″ and O sites, respectively. Depending on $t$, they crystallize in different types of crystal structures. Additionally, presence of two transition metal ions at the same site of the perovskite structure can enhance the magnetic property or tune/induce functional properties as compared with their end members. In this context, one of the well-studied series crystallizing in orthorhombic structure is RFe$_{0.5}$Cr$_{0.5}$O$_3$ (R= Tb, Dy, Ho, Er, Yb, Lu, and Tm) [2, 13-17]. Recent reports on DyFe$_{0.5}$Cr$_{0.5}$O$_3$ suggest the presence of large MCE and magnetization value enhanced by magnetoelectric coupling [14]. Also, there are evidences of enhancement of MCE and tuning of magnetoelectric coupling by rare-earth substitution and hole doping at Dy-site in DyFe$_{0.5}$Cr$_{0.5}$O$_3$ [5, 15]. Similarly, studies on HoFe$_{0.5}$Cr$_{0.5}$O$_3$ show the evidence of large MCE along with the presence of non-linear magnetodielectric behaviour [16, 17]. Magnetization studies on LuFe$_{0.5}$Cr$_{0.5}$O$_3$ [2] reveal that it undergoes magnetization reversal at compensation temperature $T_{comp}$ ~ 225 K. Recent neutron diffraction studies show that ErFe$_{0.5}$Cr$_{0.5}$O$_3$ undergoes two spin reorientation (SR) transitions near 150 K and 8 K, while YbFe$_{0.5}$Cr$_{0.5}$O$_3$ undergoes only one SR transition below 50 K [2, 13]. Generally, occurrence of magnetic phase transition is accompanied by large change in magnetization value. This is responsible for a change in isothermal magnetic entropy across phase transition, which is an important requirement to observe large magnetocaloric effect. Additionally, in such systems, mutual coupling exists between electric and magnetic ordering which can be mediated through spin-phonon interaction. Hence, in this context it may be interesting to explore these possible correlations in ErFe$_{0.5}$Cr$_{0.5}$O$_3$ and YbFe$_{0.5}$Cr$_{0.5}$O$_3$. To the best of our knowledge, such studies on these two compounds are absent in literature.

In this paper, we present our detailed investigation on ErFe$_{0.5}$Cr$_{0.5}$O$_3$ and YbFe$_{0.5}$Cr$_{0.5}$O$_3$ compounds through magnetic, dielectric and Raman spectroscopy studies. Significant enhancement of magnetic entropy change is noted in ErFe$_{0.5}$Cr$_{0.5}$O$_3$, as compared to that observed in YbFe$_{0.5}$Cr$_{0.5}$O$_3$. At high temperatures, in both the compounds, due to charge carrier hopping between Fe$^{3+}$ and Cr$^{3+}$ ions (in the presence of applied electric field), Debye-like relaxation behaviour is observed. This is confirmed through the analysis of Cole-Cole plot. Interestingly, temperature dependent Raman scattering studies divulge that there is anomalous change in the behaviour of line width and phonon frequencies near SR transitions. This deviation clearly demonstrates that spin-phonon coupling plays a crucial role in influencing the magnetic interactions between R$^{3+}$ and Cr$^{3+}$/Fe$^{3+}$ ions in these mixed metal oxides.

## 2. Experimental details

Polycrystalline compounds of ErFe$_{0.5}$Cr$_{0.5}$O$_3$ (ErFCO) and YbFe0$_{.5}$Cr$_{0.5}$O$_3$ (YbFCO) are synthesized by solid state reaction route under identical conditions as described in [3]. Structural analysis of the compounds are carried using room temperature x-ray diffraction (XRD) patterns obtained from rotating anode Rigaku Smartlab diffractometer in Bragg-Brentano geometry (Cu-Kα; λ=1.5406 Å). The Rietveld refinement of the obtained XRD patterns is performed using FullProf Suite software. Temperature and magnetic field dependent DC magnetization and AC susceptibility measurements have been performed using the Magnetic property measurement system (MPMS), Quantum Design, USA. Complex dielectric permittivity is recorded as a function of temperature by using Hioki LCR meter integrated with Physical Property Measurement System (PPMS), Quantum Design, USA with a setup from Cryonano Labs. The measurements are conducted with ac bias ~ 1V and at different frequencies (5-300 kHz). Raman spectra of the compounds are obtained at different temperatures (4-300 K) in back scattering geometry by using Horiba HR-Evolution spectrometer with 633 nm excitation laser.

## 3. Results and discussions

### 3.1. Structural properties

Fig. 1 (a) and (b) represent the room temperature XRD patterns of ErFCO and YbFCO. It can be inferred that both compounds crystallize in orthorhombic structure with Pbnm space group. The structural parameters obtained from Rietveld refinement are listed in table I. As noted from the table I, the lattice parameter of unit cell for the latter compound is less as compared to the former compound. A comparative analysis of the XRD spectra of ErFCO and YbFCO reveals that diffraction lines show gradual shift towards higher angle side in YbFCO, due to lanthanide contraction (shown in inset of Fig. 1(b)). The value of $t$ indicates that ErFCO is more distorted as compared to YbFCO.

### 3.2. Temperature and field dependent DC and AC susceptibilities studies

Temperature dependent DC magnetic susceptibility ($\chi_{DC}$) curves obtained under zero-field-cooled (ZFC), field-cooled-cooling (FCC) and field-cooled-warming (FCW) conditions at 100 Oe for ErFCO and YbFCO are shown in Fig. 2 (a) and (b) respectively. In case of ErFCO, it is observed that as temperature is decreased, bifurcation between ZFC and FCW curves is noted around 270 K ($T_{N1}$); followed by a change in slope 150 K ($T_{N2}$), which is visible in ZFC curve. Below 90 K, susceptibility value increases and attains maxima till 9.5 K followed by sudden drop in its value near 8 K ($T_{N3}$). Interestingly, weak thermal hysteresis is also observed between FCC and FCW curves. As reported previously [13], $T_{N1}$ arises due to G-type antiferromagnetic ($\Gamma_4(G_xA_yF_z)$) (AFM) ordering of the $Fe^{3+}/Cr^{3+}$ sub-lattices and $T_{N2}$ occurs because of the onset of first SR transition. During the latter transition the magnetic structure of the unit cell changes from $\Gamma_4(G_xA_yF_z)$ to $\Gamma_2(F_xC_yG_z)$. The third transition $T_{N3}$ arises due to the occurrence of $2^{nd}$ SR transition ($\Gamma_2(F_xC_yG_z) \rightarrow \Gamma_1(A_xG_yC_z)$). During this transition, there is appearance of C-type antiferromagnetic ordering of $Er^{3+}$ ions on its sub-lattice. However, in case of YbFCO, it is observed from ZFC and FCW curves that separation between them starts from ~ 280 K ($T_{N1}$). At $T_{N1}$, G-type AFM ordering of moments on $Fe^{3+}/Cr^{3+}$ sub-lattice is reported [2]. As temperature is decreased, a peak is observed near 36 K ($T_{N2}$). This observation is attributed to occurrence of first SR transition ($\Gamma_4(G_xA_yF_z) \rightarrow \Gamma_2(F_xC_yG_z)$), and is similar to that observed in ErFCO. Interestingly, below 25 K, both ZFC and FCW curves indicate towards an increment in the susceptibility value. This feature has been reported as the effect of negative thermal expansion (NTE) arising due to repulsion between magnetic moments of neighbouring transition metal ions [2]. However, such features are absent in ErFCO. Additionally, we have noticed large thermal hysteresis between FCC and FCW below $T_{N1}$ in this compound (as shown in Fig. 2(b)) which is also negligible in ErFCO. Such thermal hysteresis is not unusual and has been reported in $DyFe_{0.5}Co_{0.5}O_3$, $Dy_2FeCoO_6$ [18, 19]. This feature can be explained on the basis of interaction strength and effect of temperature on magnetic sub-lattices of both compounds. For YbFCO, in FCC curve, as the temperature decreases, the $Fe^{3+}/Cr^{3+}$ ions undergoes G-type antiferromagnetic order (along a-axis) with weak ferromagnetic component (along c-axis) near $T_{N1}$. This ordering produces an effective field on $Yb^{3+}$ ions. The Yb sub-lattice experiences a combined field which is the sum of the internal field due to Fe/Cr sub-lattice and the external applied field. Hence, the resultant magnetization is combination of magnetization from Fe/Cr and Yb sub-lattices. Near 36 K, YbFCO undergoes 1$^{st}$ SR transition, where $Fe^{3+}/Cr^{3+}$ magnetic moments change its orientation from a-axis to c-axis ($\Gamma_4(G_xA_yF_z) \rightarrow \Gamma_2(F_xC_yG_z)$). This is observed as the effect of competition between anisotropic field of Fe/Cr and their interactions with $Yb^{3+}$ ions. The anti-symmetric and the anisotropic-symmetric exchange interaction produce an effective field which forces the spin up $Cr^{3+}/Fe^{3+}$ ions in the direction perpendicular to previous spin arrangement and an effective field for the down spins $Cr^{3+}$ in the direction opposite to the above one. As the temperature is further lowered, interaction energy of $Yb^{3+}$ ions with the $Fe^{3+}/Cr^{3+}$ dominates over anisotropic field of $Fe^{3+}/Cr^{3+}$, rotating the $Cr^{3+}/Fe^{3+}$ ions along c-axis. While in FCW, when the compound is warmed, spins may not fully re-orient from $\Gamma_2(F_xC_yG_z) \rightarrow \Gamma_4(G_xA_yF_z)$. This is observed as $Cr^{3+}/Fe^{3+}$-$Yb^{3+}$ magnetic interaction do not produce strong enough fields to rotate the $Fe^{3+}/Cr^{3+}$ spins. Therefore, we have observed a large difference between FCC and FCW magnetization curves in YbFCO compound. However, in ErFCO, $Er^{3+}$-$Cr^{3+}/Fe^{3+}$ magnetic interaction produces strong effective field to rotate $Fe^{3+}/Cr^{3+}$ spins, leading to negligible difference between magnetization value obtained in FCC and FCW measurements.

To get a better understanding about the complex magnetic behaviour of both compounds, we have also measured isothermal magnetization (*M*) as a function of magnetic field (*H*) at different temperatures (2-150 K). Fig. 2 (c) and (d) show the corresponding *M* (*H*) curves obtained at different temperatures (2, 40 and 150 K) for ErFCO and YbFCO, respectively. In both compounds, weak magnetic hysteresis is noted below 15 kOe and magnetization does not saturate at high fields. This type of behaviour reflects the existence of canted AFM state along with FM correlations. For ErFCO, at 2 K, small change in the curvature of virgin curve is observed near 6.5 kOe suggesting the presence of metamagnetic transition (MM) (as seen in inset of Fig. 2 (c)). However, this magnetic feature is suppressed at higher temperatures (40 and 150 K) as shown in Fig. 2 (c). This metamagnetic transition is associated with the $\Gamma_1(A_xG_yC_z)$ magnetic structure. However, in case of YbFCO, at 2 K, no signature of metamagnetic transition is visible in $M$ ($H$) curves. Similar to ErFCO, in this compound we have also noted the non-linear behaviour of magnetization with applied field along with presence of small hysteresis at 40 and 150 K.

In order to probe whether the observed thermomagnetic irreversibility is arising due to the presence of any glassy phase, we have performed AC susceptibility measurements in the temperature range 2-300 K at different frequencies (13-931 Hz) at $H_{AC}$ =1 Oe. Fig. 3 (a) and (b) illustrate the temperature dependent in-phase ($\chi'$) part of AC susceptibility at different frequencies of ErFCO and YbFCO, respectively. In ErFCO, $\chi'$ exhibits magnetic anomalies near $T_{N2}$ and $T_{N3}$ which is analogous to that observed in $\chi_{DC}$ curve. However, we have not observed any frequency dependent shift in temperature around these transitions. For YbFCO, $\chi'$ shows an anomaly near $T_{N2}$. Frequency dependence of $\chi'$ is not noted in YbFCO. We have not shown out-of-phase part ($\chi''$) as its magnitude is comparable with the error bar of data measurement, for both the compounds. Additionally, in both compounds, no anomaly is visible near $T_{N1}$ transition as there is no significant change noted in magnetization value with applied field near this transition. This absence of frequency dependence in $\chi'$ in 2-300 K temperature regime rules out the presence of any glassy dynamics in both compounds.

From the above results, it can be said that interaction between rare-earth ions and $Fe^{3+}/Cr^{3+}$ sub-lattice plays a significant role in deciding the behaviour of temperature and magnetic field dependent magnetization curves of ErFCO and YbFCO. The observed SR transitions in these compounds are triggered by both anisotropic-symmetric exchange and anti-symmetric Dzyaloshinskii- Moriya (DM) interactions between $Fe^{3+}/Cr^{3+}$ and $R^{3+}$ ions. Therefore, transition temperatures are conditioned by the combination of cations and their magnetic interactions with rare-earth ions [2, 13]. In the present case, we have seen that the $Cr^{3+}$-$Fe^{3+}$/$Fe^{3+}$-$Fe^{3+}$/$Cr^{3+}$-$Cr^{3+}$ interactions are dominant above 200 K, and this interaction is responsible for the AFM type ordering of $Fe^{3+}/Cr^{3+}$ magnetic sub-lattices near 270 K and 280 K in ErFCO and YbFCO, respectively. However, in ErFCO with reduction in temperature, it is found that the $Fe^{3+}/Cr^{3+}$ sub-lattice undergoes 1st SR transition near 150 K, which can be due to dominance of the $Er^{3+}$-$Cr^{3+}$/$Fe^{3+}$ magnetic exchange interaction over $Cr^{3+}$-$Fe^{3+}$/$Fe^{3+}$-$Fe^{3+}$/$Cr^{3+}$-$Cr^{3+}$ interactions. At 8 K, the $Er^{3+}$ paramagnetic ions order cooperatively resulting in a change of alignment of $Fe^{3+}/Cr^{3+}$ sub-lattice in y-direction (2nd SR transition) [13]. In contrast to this, in YbFCO, it can be seen that below 280 K, the interaction strength between $Yb^{3+}$ and $Fe^{3+}/Cr^{3+}$ ions is weaker compared to $Cr^{3+}$-$Fe^{3+}$/$Fe^{3+}$-$Fe^{3+}$/$Cr^{3+}$-$Cr^{3+}$ exchange interaction [2]. As the temperature is reduced, the strength of this interaction increases leading to 1$^{st}$ SR transition near 36 K. Since, $Yb^{3+}$-$Yb^{3+}$ paramagnetic ions do not order magnetically on $Fe^{3+}/Cr^{3+}$ sub-lattice, only one type of SR transition is seen in this compound.

Also, in order to identify the order of transition in the vicinity of $T_{N3}$ and $T_{N2}$ in ErFCO and YbFCO, respectively, $H/M$ vs $M^2$ plots are obtained from virgin curves of isothermal magnetization. Figure 4 (a) and (b) shows the $H/M$ Vs $M^2$ plot for ErFCO and YbFCO at selected temperatures, respectively. This Arrott plots exhibit negative and positive slope for first and second order phase transition. For ErFCO, positive slope change is observed near $T_{N3}$ indicating the second order nature of this transition. It is interesting to note that there is crossover from negative (6 K) to positive slope (8 K) which supports the possibility of occurrence of MM transition in this temperature range. Similarly, in case of YbFCO, a positive slope change is noted around $T_{N2}$ transition which also suggests the occurrence of second order transition in this compound.

### 3.3. Magnetocaloric effect

Literature reports [14, 15, 17, and 20] on mixed metal oxides suggest the presence of significant MCE across the magnetic phase transitions, where significant change in magnetization occurs. As observed from Fig. 2 (a) and (b), the magnitude of magnetization changes abruptly in the vicinity of $T_{N3}$ and $T_{N2}$ in ErFCO and YbFCO, respectively. Hence, it is of interest to investigate the temperature dependent behaviour of MCE of ErFCO and YbFCO compounds around these transition temperatures. Here, MCE is calculated in terms of isothermal magnetic entropy change ($\Delta S_M$) produced by the changes in magnetic field. In order to probe the same, we have measured $M(H)$ isotherms in the vicinity of above mentioned transitions in both compounds. Using the virgin curves of obtained $M(H)$, we have calculated $\Delta S_M$ using the following equation [21]:

$$\Delta S_M = \sum \frac{M_{n+1} - M_n}{T_{n+1} - T_n} \Delta H_n \quad \text{............} (1)$$

where $M_n$ and $M_{n+1}$ are the magnetization values obtained at field $H_n$ at temperature $T_n$ and $T_{n+1}$ respectively. Fig. 4 (c) and (d) show the temperature dependent $\Delta S_M$ curves for different applied field of ErFCO and YbFCO respectively. For ErFCO, it is observed that $\Delta S_M$ attains maximum value of ~ -12.4 J/kg-K near 8 K for $\Delta H$ = 50 kOe and peak temperature of $\Delta S_M$ matches well with $T_{N3}$. However, in case of YbFCO, the value of $\Delta S_M$ decreases to -3.2 J/kg-K for same $\Delta H$ near 7.5 K. Here, it is interesting to note that peak temperature of $\Delta S_M$ is different from $T_{N2}$.

As mentioned before, abrupt change in magnetization value give rise to considerable amount of magnetic entropy change, which results in significant MCE. Additionally, magnetic ordering of rare-earth ions also plays an important role in deciding the value of $\Delta S_M$ in these types of compounds. For ErFCO it can be seen that $Er^{3+}$ ions magnetically order on $Fe^{3+}/Cr^{3+}$ sub-lattice below 8 K and significant $\Delta S_M$ is also noted near this ordering. Whereas, due to absence of ordering of $Yb^{3+}$ ions on $Fe^{3+}/Cr^{3+}$ lattice, a low value of $\Delta S_M$ is observed in YbFCO, as compared to ErFCO. Interestingly, the later compound has higher value of $\Delta S_M$ at $\Delta H$ = 50 kOe as compared to other transition metal oxides such as $DyFe_{0.5}Cr_{0.5}O_3$, $SmCr_{0.85}Mn_{0.15}O_3$, rare-earth substituted $DyCrO_3$ [14, 22, 23] etc. Additionally, as noted in YbFCO, $\Delta S_M$ maximum does not coincide with $T_{N2}$ but it is around 7.5 K. This type of deviation is not uncommon and has been reported in literature [5, 24]. In YbFCO, NTE is observed in the low temperature regime and such effect was also reported in $Dy_{1-x}Ho_xMnO_3$ [24]. In YbFCO, NTE leads to distortion of $FeO_6/CrO_6$ polyhedron resulting in disappearance of $A_{g(1)}$ and $A_{g(2)}$ Raman modes (further discussed in section 3.5). This changes the crystal field significantly. This can possibly affect the $Fe^{3+}/Cr^{3+}$ interaction strength which is dominant on sub-lattice (as $Yb^{3+}$ ions ordering is absent till 4 K); leading to the observed deviation in the temperature of the peak of $\Delta S_M$.

Relative cooling power (RCP) is a parameter which is calculated to check the efficiency of any material that can be used as refrigerant material. RCP gives the measure of amount of heat transferred between cold and hot end in an ideal refrigeration cycle. In general, it is defined as the product of maximum of $\Delta S_M$ ($\Delta S_M^{max}$) and full width at half maximum of the peak of $\Delta S_M$ ($\Delta T_{FWHM}$) [25]. But, in mixed metal oxides, temperature dependent behaviour of $\Delta S_M$ is not symmetric in nature. It is calculated as [23]

$$\text{RCP} = \int_{T_h}^{T_c} | \Delta S_M (T, \Delta H) | \, dT \quad \ldots\ldots\ldots\ldots (2)$$

where $T_C$ and $T_h$ are the temperature of cold and hot end of thermodynamic cycle respectively. In case of ErFCO, it is found that value of RCP is ~ 221.14 J/kg at maximum applied field ($\Delta H$=50 kOe), whereas it is ~48.51 J/kg in case of YbFCO under same applied field. It can be seen that the value of RCP for YbFCO is much less as compared with ErFCO and other mixed metal oxides. Here, it can be said that ErFCO shows good magnetocaloric properties in comparison with YbFCO in the cryogenic temperature regime and presence of a magnetic rare-earth ions can significantly enhance the magnetocaloric properties of these mixed metal oxides.

In order to get better insight about the MCE properties of both compounds, we also have analyzed the field dependent $\Delta S_M$ behaviour at selected temperature with power law [3, 26-28] i.e. $\Delta S_M \sim H^n$, where $n$ is the exponent which is related to magnetic state of compounds. We have fitted the field response of $\Delta S_M$ with power law above peak temperature. Fig. 4 (e) and (f) represent the curves of $\Delta S_M$ vs $H$ at selected temperatures for ErFCO and YbFCO respectively. For ErFCO, it is found out the $\Delta S_M$ follows power law and obtained values of $n$ lies between 1.3-1.8. Similarly, in YbFCO, $\Delta S_M$ also follows power law but obtained values of $n$ (1.8-1.9) is quite higher than those obtained in case of ErFCO. However, it is found that the values obtained in both cases are less than 2 (ideal AFM system) [28]. This lower value of $n$ reflects the presence of FM correlations in AFM state which supports our conclusion drawn from previous sections.

### 3.4. Dielectric properties

Compounds belonging to the mixed metal oxide family are found to exhibit ferroelectric ordering near magnetic phase transitions, which suggests that there is some correlation between magnetic and electric ordering in these compounds [5, 14-16]. Thus, in order to probe whether there is existence of similar dependence in ErFCO and YbFCO compounds, we have further measured the in-phase ($\varepsilon'$) and out-of-phase ($\varepsilon''$) part of the dielectric permittivity at different frequencies (5-300 kHz) in the 5-300 K temperature range. Fig. 5 (a)-(d) illustrate the temperature response of the dielectric constant $\varepsilon'$ and dielectric loss factor tan$\delta$ (tan$\delta = \varepsilon''/ \varepsilon'$) at different frequencies of both compounds. In ErFCO, it is found that $\varepsilon'$ decreases with increase in frequency and broad anomaly can be clearly seen around 8 K (right inset of Fig. 5 (b)). This anomaly can be due to the onset of ferroelectric (FE) transition ($T_{FE\_Er} \sim 8$ K). Interestingly, in $\chi_{DC}$ ($T$), we have also noted a magnetic transition near this temperature, which suggests that there can be some correlation between magnetic and electric order in this compound. The reduction in magnitude of $\varepsilon'$ with frequency can be due to reduced space charge polarization effect with frequency, while, sharp step-like increase is noted above 150 K but it does not attain maxima till 300 K. Here, this type of behaviour of dielectric dispersion can be attributed to Maxwell-Wagner type of interfacial polarization [29]. However, tanδ increases with increment in temperature and attains a maximum near 261 K. This behaviour reflects an increment in mobility of charge carriers and signifies the accumulation of charge at grain boundaries at high temperature. It is also noticed that this maxima shows strong frequency dependent shift towards high temperature side. Similarly in YbFCO, in temperature response of ε′ we have observed an anomaly near 36 K ($T_{FE\_Yb}$), which can be due to onset of ferroelectric transition in this compound (right inset of Fig. 5 (d)). Along with it, a step-like increase in value of ε′ is noted above 120 K, while tanδ exhibits frequency dependent maxima near 218 K. However, in this compound, the observed value of ε′ is very large compared to ErFCO. Interestingly, both compounds show ferroelectric transition near magnetic ordering temperature, which indicates there is significant coupling between the magnetic and dielectric properties.

The shift in maxima towards higher temperature side with frequency in both compounds indicates the glassy behaviour of electric dipoles in high temperature region. This behaviour can be further analysed using either Arrhenius law or Vogel-Fulcher law. In order to analysis this variation of peak maxima, we have fitted this response with Arrhenius law (as it gives the best fit as compared to Vogel-Fulcher law) which is stated as [30]

$$\tau = \tau_0 e^{\frac{E_a}{k_B T}} \ldots\ldots\ldots\ldots (3)$$

where $\tau_o$ is the pre-exponential factor and $E_a$ is the activation energy. Left insets of Fig. 5 (b) and (d) represent the log τ vs. $1/T_{peak}$ plot for both compounds. It can be noted that in both compounds, relaxation process can be described by Arrhenius law. The obtained values of $E_a$ for ErFCO and YbFCO are 0.41±0.04 eV and 0.14±0.004 eV respectively. These values are comparable with the values reported in other oxides, where the relaxation mechanism is attributed to charge carrier hopping between transition metal ions [31, 32]. In the present scenario, dielectric relaxation can arise from two independent mechanisms: i) Hopping of charge carriers between $Cr^{3+}$ and $Fe^{3+}$ in the presence of electric field, which give rise to Debye type of relaxation; ii) Accumulation of charge carriers between regions in the sample i.e. grain boundaries, this type of relaxation is Maxwell-Wagner relaxation. Signatures of second mechanism are reflected in the behaviour of ε′ as discussed above. In order to verify the role of first mechanism in ErFCO and YbFCO, the complex impedance spectrum in the form of Cole-Cole (ε″ vs. ε′) plot obtained near $T_{peak}$ is analyzed using equation [33]

$$\varepsilon^* = \varepsilon' - j\varepsilon'' = (\varepsilon_\infty + (\frac{\varepsilon_s - \varepsilon_\infty}{1+(j\omega\tau)^{1-\alpha}}))\ldots\ldots\ldots\ldots (4)$$

where, $\varepsilon_\infty$ is the high frequency limit of the permittivity, $\varepsilon_s$- $\varepsilon_\infty$ is the dielectric strength, $\tau$ is the mean relaxation time ($\tau = \omega^{-1}$; $\omega = 2\pi\nu$) and $\alpha$ is the measure of distribution of relaxation time, it is equal to zero in case of mono-dispersive Debye relaxation process. Cole-Cole plot obtained at peak temperatures for both compounds are shown in inset of Fig. 5 (a) and (c). For ErFCO and YbFCO, the obtained values of $\alpha$ at different temperatures above and below $T_{peak}$ are found near to zero, suggesting Debye type of dielectric relaxation in high temperature region. Also, in ErFCO and YbFCO, there is existence of a single semicircle in both compounds. This type of single relaxation behaviour is observed in other systems like Co-Ni-Li ferrite [34], $Al^{3+}$ Ni-Zn ferrite [35] and it indicates that contribution of grains is dominant over that of grain boundaries in both compounds.

### 3.5. Raman spectroscopic studies

Rajeswaran et al. [36] have reported occurrence of electric polar order near the magnetic ordering temperature of Cr in $RCrO_3$ (R = rare-earth ions), where it has been suggested that interaction between $Cr^{3+}$ and $R^{3+}$ is responsible for observed polarization. Similarly, in ErFCO and YbFCO we have noted that $T_{FR\_Er}$ and $T_{FR\_Yb}$ are near to $T_{N3}$ and $T_{N2}$ respectively. In this type of compounds, this interplay between magnetic and electric ordering can be mediated through spin-phonon coupling. Thus, probing the local structure can give a better insight of the multiferroic behaviour of these two compounds.

Raman spectroscopy is an ideal technique to explore the local structural changes due to magnetic ordering in compounds. In order to study this, we have measured temperature dependent Raman scattering spectra in temperature range 4-300 K. Fig. 6 (a)-(d) represents the Raman spectrum for both compounds in the frequency range of 95-715 cm$^{-1}$ at 4 and 300 K respectively. It has been previously observed that both compounds crystallize in an orthorhombic (Pbnm) structure. An ideal perovskite have a cubic structure. One can obtain the orthorhombic structure from cubic one by either by: (a) rotation or tilting of $CrO_6/FeO_6$ octahedra around [101] or [010] (for these mixed metal oxides), (b) displacement of rare-earth ions. These structural distortions lead to lowering of crystal symmetry which in turn activates the Raman modes. According to group theory, there are 24 raman active modes possible for orthorhombic structure with 4 formula units/unit cell ($7A_g + 5B_{1g} + 7 B_{2g} + 5B_{3g}$) [37]. However, in case of ErFCO and YbFCO, at 300 K and 4 K, we have observed less than 24 Raman active phonon modes. These modes are assigned as $A_g$, $B_{1g}$, $B_{2g}$ and $B_{3g}$ [38, 39] and are shown in Fig. 6 (a)-(d).

From Fig. 6 (a) and (c), at 300 K, Raman spectra of both compounds are similar. There is no significant difference between the Raman active modes of these two compounds because in high temperature regime $R^{3+}$-$R^{3+}$ interactions are insignificant. However, there is a difference between the Raman spectra obtained at 300 K and 4 K. In ErFCO, at 4 K additional modes $A_{g(5)}$ and $B_{2g(1)}/B_{2g(7)}$ are present. These modes may arise due to change in bond length/angle of R-O-Cr/Fe and titling of $FeO_6/CrO_6$ octahedral as temperature is decreased below $T_{N3}$. In YbFCO, as the temperature is decreased, there is disappearance of $A_{g(1)}$ and $A_{g(2)}$ phonon modes at 4 K. These modes completely disappear below 20 K, where we have earlier observed NTE. This disappearance of the modes might be due to the fact that owing to the expansion of unit cell there is a decrement in energy of these modes [40]. Further the temperature dependence of Raman modes of ErFCO and YbFCO in different wave number regime is explained below in detail.

In low wave number region (below 200 cm$^{-1}$), the observed Raman modes depends upon the rare-earth ion vibrations. So, according to harmonic oscillator approximation as the mass increases from Er to Yb by about 3%, so the vibration frequencies ($A_{g(1)}$, $A_{g(2)}$ and $B_{2g(1)}$) associated with rare-earth ion decreases a little (as shown in Fig. 7 (a)). The mid-wave number region (200-380 cm$^{-1}$) is characterized by the presence of two doublets which are assigned as $B_{1g(1)}/A_{g(3)}$ and $B_{2g(2)}$ and $A_{g(4)}$ modes. It is observed that $B_{1g(1)}$ and $A_{g(3)}$ modes overlap (and hence treated as one). All these modes are generally affected by the ionic radii of rare-earth ions. They show shift towards low wave number with increasing ionic radii of rare-earth. As the ionic radii of $Er^{3+}$ is more than $Yb^{3+}$, it can be seen that $A_{g(4)}$ mode shifts towards high wave number in case of YbFCO compared to ErFCO (Fig. 8 (a)). However, in high wave number regime (380-715 cm$^{-1}$), we have also noted a shift in the $B_{3g(5)}$ mode in YbFCO with respect to that in ErFCO; towards high wave number side (from 674.728 to 676.094 cm$^{-1}$). This mode is associated with anti-symmetric or symmetric stretching of $CrO_6/FeO_6$ octahedra [38, 39], which can be due to orbital mediated electron phonon coupling.

Temperature dependencies of phonon frequencies and corresponding line widths of $A_{g(4)}$, $B_{2g(1)}$ and $B_{3g(5)}$ are shown in Figs. 7 (b), 7(c), 8 (b), 8 (c), 9 (a) and 9 (b), as these modes are expected to be affected by R-O vibrations and stretching of $FeO_6/CrO_6$ octahedra respectively. In all these three modes, abrupt change in the phonon frequency is noted near $T_{N3}$ and $T_{N2}$ for ErFCO and YbFCO, respectively. Similar changes were previously reported in other compounds near magnetic and ferroelectric transitions [14, 41]. Interestingly, we have also noted anomalous behaviour in the line width (FWHM) of above-mentioned modes. Since the line width is related to phonon life time and does not depend on lattice volume change, therefore sudden increase in the linewidth of $A_{g(4)}$ and $B_{2g(1)}$ near $T_{N3}$ and $T_{N2}$ as temperature is increased indicates decrease in phonon lifetime due to the existence of spin-phonon coupling in both compounds and is associated to the renormalization of the phonons induced by the magnetic ordering [42-45]. The changes in phonon lifetime also suggest the presence of disorder in the compounds due to magnetic order. This type of disorder can be influenced near magnetic ordering due to titling or rotation of octahedral, ordering of Fe/Cr ions in oxygen octahedral or due to displacement of the rare-earth ion [46]. Additionally, from Figs. 7 (b) and (c), below $T_{N3}$ in ErFCO and $T_{N2}$ in YbFCO, a contrasting behaviour is observed for $B_{2g(1)}$ mode. An anomalous softening of the mode is observed for YbFCO compound whereas for ErFCO, a hardening of this mode is noted. This type of different behaviour can be ascribed to the presence of different type of magnetic interactions between $R^{3+}$, $Cr^{3+}$ and $Fe^{3+}$ ions [37].

As observed from Figs. 7 (b), 7 (c), 8 (b), 8 (c), 9 (a) and 9 (b); $A_{g(4)}$, $B_{2g(1)}$ and $B_{3g(5)}$ show deviation near $T_{N3}$ and $T_{N2}$ in case of ErFCO and YbFCO respectively. $A_{g(4)}$ and $B_{2g(1)}$ modes associated with $R^{3+}$ ions ($Er^{3+}/Yb^{3+}$) vibrations exhibiting deviation near $T_{N3}$ and $T_{N2}$ indicate that the above mentioned magnetic transitions have a close relation with $R^{3+}$ ions. Also, it is observed that there is change in phonon frequencies along with decrease in phonon lifetime in modes involving both $CrO_6/FeO_6$ octahedra and $R^{3+}$ magnetic ions near $T_{FE\_Er}$ and $T_{FE\_Yb}$. These results indicate that there is presence of phonon mediated magnetic interactions between $R^{3+}$ and $Cr^{3+}/Fe^{3+}$ ions. These interactions and softening of R ion modes suggest that the occurrence of ferroelectric behaviour in these compounds can be related to the displacement of rare earth ions. Similar results are reported in other systems like $RCrO_3$ [36] and $YFe_{1-x}Mn_xO_3$ [47]. Hence it can be said that the temperature dependent Raman results indicate the anomalous behaviour of phonon modes. This behaviour is related to vibration of rare-earth atoms and stretching of $CrO_6/FeO_6$ octahedra and it is noted near spin-reorientation and ferroelectric transitions of compounds. This suggests that the ferroelectricity in ErFCO and YbFCO is associated with spin-phonon coupling with respect to both Er/Yb and Cr/Fe ions.

## 4. Summary

In summary, we present a systematic investigation of the magnetic, magnetocaloric dielectric and Raman spectroscopy studies of ErFe$_{0.5}$Cr$_{0.5}$O$_3$ and YbFe$_{0.5}$Cr$_{0.5}$O$_3$. Two transitions are noted for the latter compound in contrast to the three transitions observed in the former compound. Debye-like relaxation of electric dipoles due to charge carrier hopping between Fe$^{3+}$ and Cr$^{3+}$ ions in the high temperature regime is noted in both the compounds. Large value of magnetocaloric parameters of ErFe$_{0.5}$Cr$_{0.5}$O$_3$ as compared to YbFe$_{0.5}$Cr$_{0.5}$O$_3$, make it suitable for magnetic refrigerant applications in cryogenic temperature regime. Temperature dependent anomalous behaviour of phonon modes reveals the presence of spin-phonon coupling across both spin-reorientation and ferroelectric transitions of these compounds.

**Acknowledgements**

The authors acknowledge the experimental facilities of IIT Mandi. KM acknowledges financial support from DST-SERB project EMR/2016/00682.

Table 1 Structural parameters obtained from Rietveld Refinement of XRD data of the compounds: Lattice parameters and tolerance factor (*t*).

| Parameters | ErFCO | YbFCO |
|---|---|---|
| a (Å) | 5.246(1) | 5.212(7) |
| b (Å) | 5.550(1) | 5.528(4) |
| c (Å) | 7.559(1) | 7.525(9) |
| V (Å$^3$) | 220.13(9) | 216.88(2) |
| $\chi^2$ | 2.98 | 2.71 |
| *t* | 0.869 | 0.854 |

**Figures-**

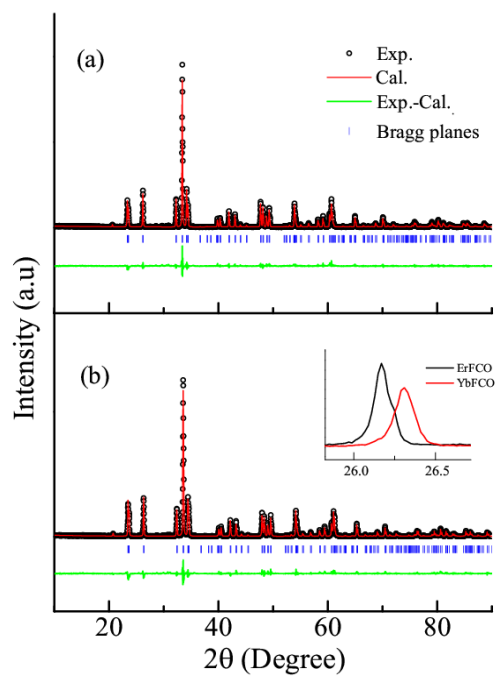

Figure 1. Room temperature XRD patterns: Rietveld analysis results for (a) ErFCO (b) YbFCO. Inset: Magnified view of XRD spectra of both to show peak shift due to presence of Yb and Er at R site.

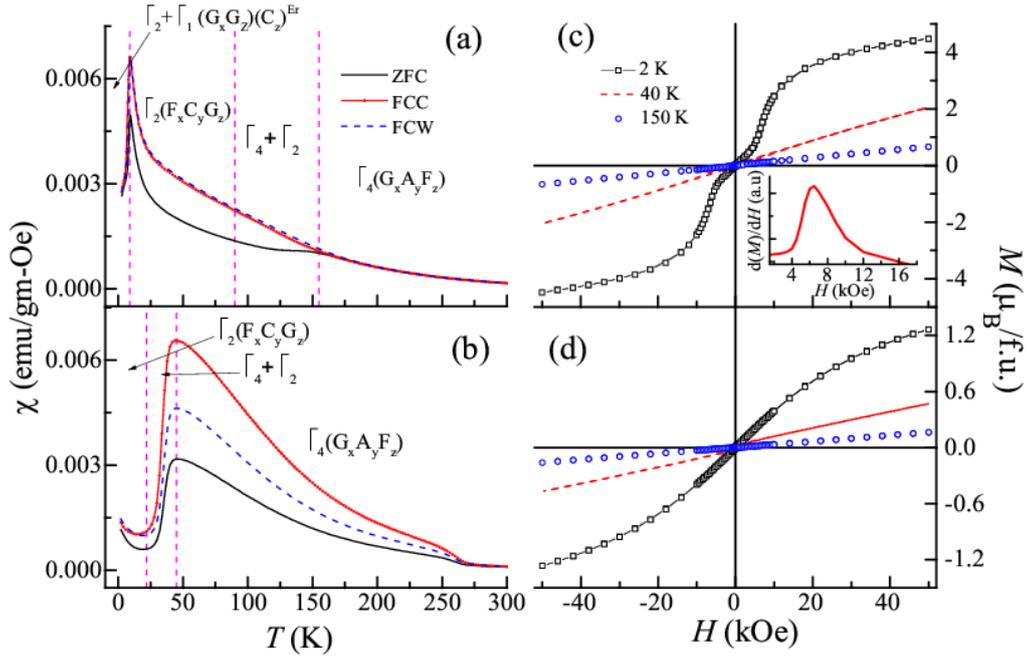

Figure 2. DC susceptibility curves of (a) ErFCO and (b) YbFCO obtained under ZFC, FCC and FCW conditions at $H = 100$ Oe in the temperature range 2-300 K. Dotted lines represent transition temperatures. Magnetic field response of magnetization of (c) ErFCO Inset: d($M$)/d$H$ vs $H$ at 2 K in the field range 1.5-18 kOe and (d) YbFCO at different temperatures.

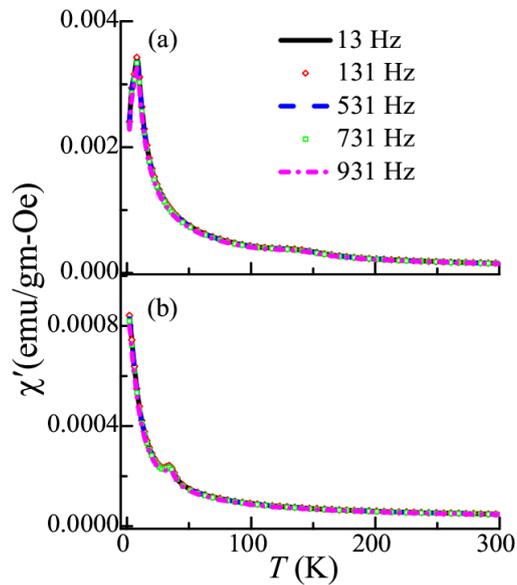

Figure 3. Temperature response of $\chi'$ part of AC susceptibility at different frequencies 13-931 Hz of (a) ErFCO and (b) YbFCO in the temperature range 2-300 K.

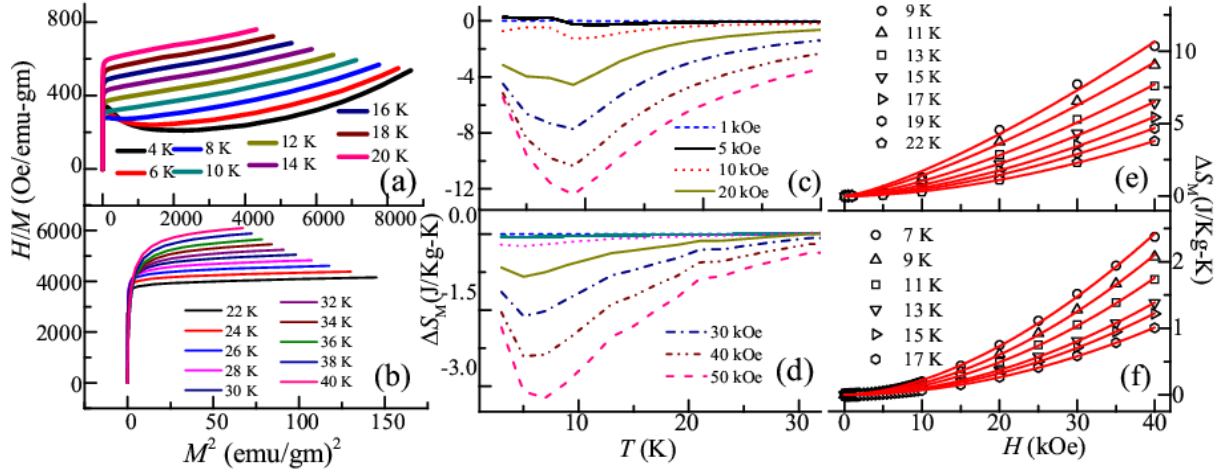

Figure 4. First panel: $H/M$ vs $M^2$ plots of (a) ErFCO and (b) YbFCO at different temperatures across $T_{N3}$ and $T_{N2}$ transitions respectively. Second panel: Temperature response of $\Delta S_M$ of (c) ErFCO (d) YbFCO at different $\Delta H$. Third panel: Magnetic field response of $\Delta S_M$ at selected temperatures for (e) ErFCO and (f) YbFCO. The solid red lines through the curves represent the power law fitting.

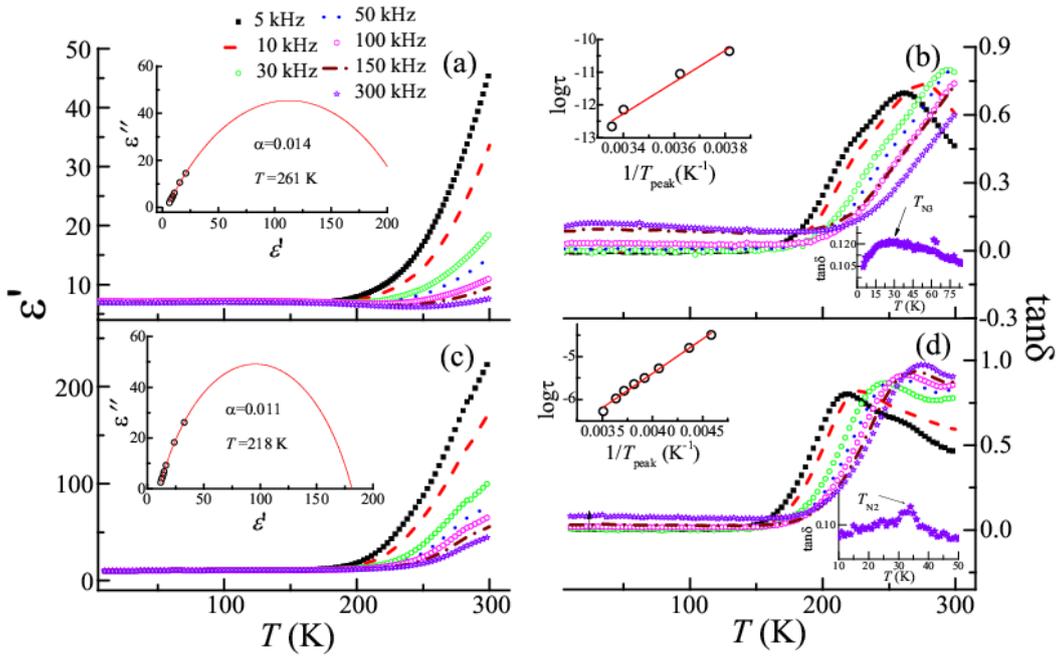

Figure 5. Left panel: Temperature ($T$) dependence of real part of dielectric permittivity ($\varepsilon'$) of (a) ErFCO and (c) YbFCO measured at different frequencies in the temperature range 5-300 K. Insets: Cole-Cole plot fitted with eqn. 4 at $T_{peak}$ for respective compounds. Right panel: Temperature dependent dielectric loss plot for (b) ErFCO and (d) YbFCO measured at different frequencies in the temperature range 5-300 K. Left Insets: log τ vs. $1/T_{peak}$ plot fitted with eqn. 3; Right Insets: Magnified view of tanδ vs. T plot at 300 kHz to show FE transitions.

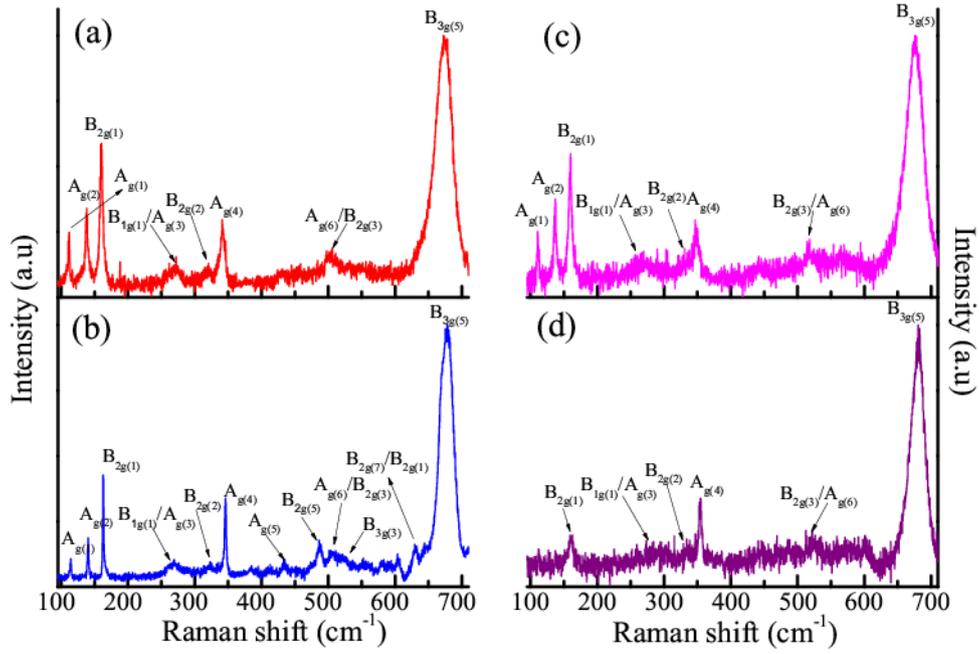

Figure 6. Raman spectra of ErFCO at (a) 300 K (b) 4 K; Raman spectra of YbFCO at (c) 300 K (d) 4 K in the range 95-715 cm$^{-1}$.

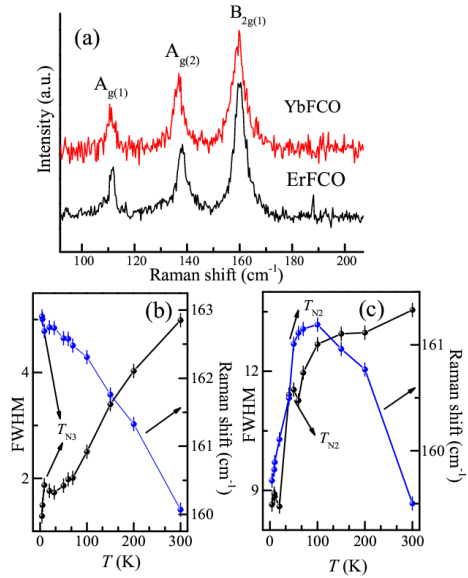

Figure 7. (a) Raman spectra of ErFCO and YbFCO at 300 K showing shift in $A_{g(1)}$, $A_{g(2)}$ and $B_{2g(1)}$ phonon modes towards high wave number side. Temperature dependence of phonon wavenumber (left) and FWHM (right) of $B_{2g(1)}$ modes of (b) ErFCO and (c) YbFCO

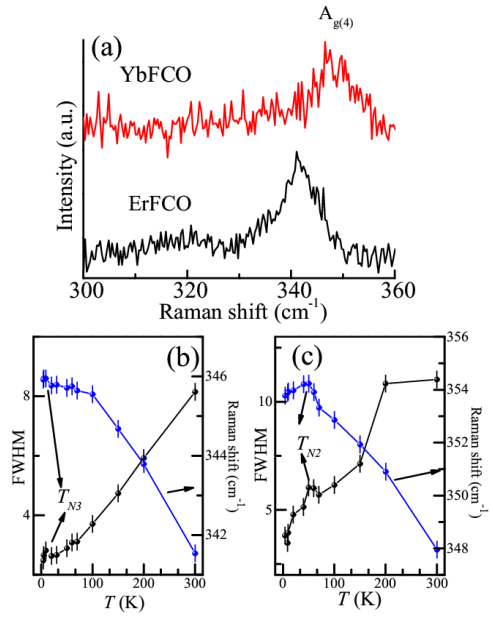

Figure 8. (a) Raman spectra of ErFCO and YbFCO at 300 K showing shift in $A_{g(4)}$ phonon modes towards high wave number side. Temperature dependence of phonon wavenumber (left) and FWHM (right) of $A_{g(4)}$ modes of (b) ErFCO and (c) YbFCO

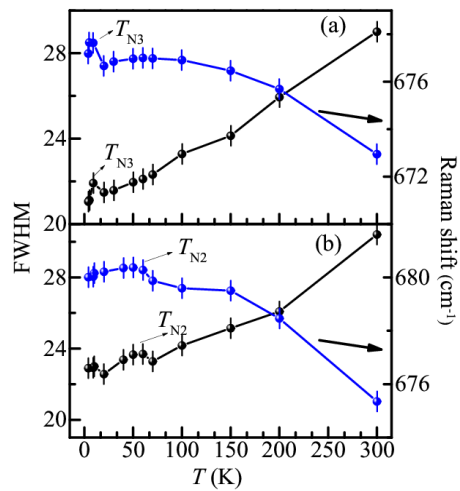

Figure 9. Temperature dependence of phonon wavenumber (left) and FWHM (right) of $B_{3g(5)}$ modes of (b) ErFCO and (c) YbFCO